\documentclass[envcountsect]{svmult}

\usepackage{mathptmx}       
\usepackage{helvet}         
\usepackage{courier}        
\usepackage{type1cm}        
\usepackage{makeidx}         
\usepackage{graphicx}        
\usepackage{multicol}        
\usepackage[bottom]{footmisc}
\newcommand\apj{\textit{ApJ}}%
\newcommand\apjl{\textit{ApJ}}%
\newcommand\apjs{\textit{ApJS}}%
\newcommand\aap{\textit{A\&A}}%
\makeindex             

\begin{document}

\title*{Dust Continuum Observations of Protostars: Constraining
  Properties with Simulations}
\author{Stella S.~R.~Offner} 
\institute{Stella S.~R.~Offner \at Harvard-Smithsonian Center for
  Astrophysics, 60 Garden St. Cambridge, MA 02138, \email{soffner@cfa.harvard.edu}}

\maketitle


\abstract{The properties of unresolved protostars and their local environment (e.g., 
disk, envelope and outflow characteristics) are frequently inferred
from spectral energy distributions (SEDs) through comparison with
idealized model SEDs. However, if it is not possible to  
image a source and its environment directly, it is
difficult to constrain and evaluate the accuracy of 
these derived properties. In this proceeding, I present a brief
overview of the reliability of SED modeling by analyzing dust continuum
synthetic observations of realistic
simulations.} 

\section{Introduction}
\label{sec:1}

Forming stars may be heavily obscured by their natal dust and gas,
which inhibits direct imaging and causes source
radiation to be reprocessed from shorter to longer wavelengths. The
details of the multi-wavelength emission, i.e., the spectral energy
distribution (SED), thus provide important indirect clues about the
protostellar properties and environment. For example, the absence of
$\le10~\mu$m emission usually signifies a very young source with a
dense gas envelope; low or non-existent millimeter emission indicates
a more evolved source, which has accreted or dispelled its envelope
(e.g., [2]). 

The information implicit in the reprocessed emission is commonly
extracted by comparing the observed SED with idealized models of the
protostellar source and gas distribution 
that are post-processed with a radiative transfer code to produce SEDs. Input models
that reproduce the observed SED provide good candidate representations of
the underlying source properties. These can provide a wealth of
physical details (e.g., source mass, disk mass and radius, envelope
density and radius, outflow cavity size, inclination) that would
otherwise be impossible to obtain with observations. However, a number
  of caveats complicate this technique, including degeneracy between
parameters, adoption of symmetry, and assumption that the observed
SED 
represents a single source rather than a multiple system or even a
small cluster [3].

Using simulations, which have completely known source and gas information, it is possible to
 assess the accuracy of this method applied to unresolved observed
 sources. In this proceeding, we summarize the results of [6], who
 present  
a comparison between the true properties of sources within simulations of a
turbulent, star formatting cloud and the properties inferred from
synthetic SEDs.


\section{Methods}
\label{sec:2}
\begin{figure}[!h]
\includegraphics[scale=.4]{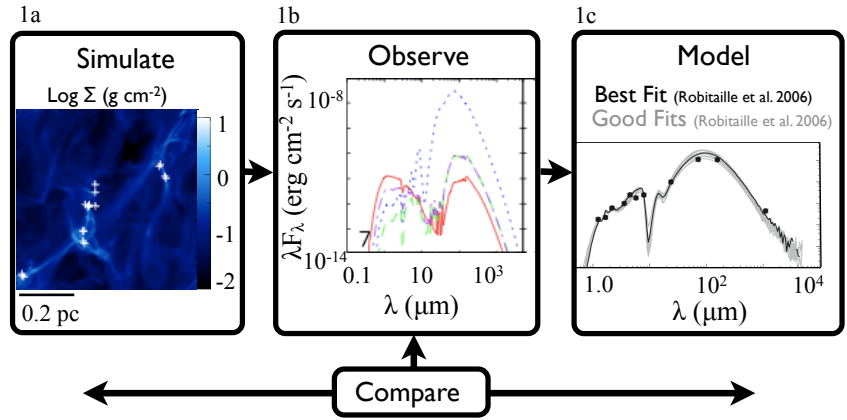}
%
%
\vspace{-0.75in}
\caption{Schematic of our process to assess the accuracy of properties
  inferred from SEDs. Panel 1a shows the log column density of the simulation 60
  kyr after the first star forms; protostar locations are marked with
  white crosses. Panel 1b shows the SED of a source observed at four inclinations by
  the Hyperion code. Panel 1c shows a synthetic SED (black points) fit with
  SED models from the [8] model grid; the best-fit
  is indicated by the black line and good-fit cases are overlaid in gray.} 
\label{scheme}
\end{figure}

To perform the comparison, we follow three main steps (see Figure
\ref{scheme}). First, we use the ORION adaptive mesh refinement
(AMR) code to simulate a 0.65 pc long turbulent cloud including self-gravity, 
radiation in the flux-limited diffusion approximation, and star
particles inserted in regions of the flow exceeding the maximum
resolution. The star particles are endowed with a sub-grid model for
protostellar evolution and mass outflow
launching based upon [4] (see [5] and [1] for implementation details). Both aspects are
coupled to the instantaneous protostellar mass and accretion rate, which are modeled
self-consistently through gas accretion from the AMR grid. For computational
efficiency, we first perform the calculation with 200 AU cell
resolution and then ``zoom'' in to 4 AU resolution by restarting and adding
additional refinement at selected outputs.

Next, we ``observe'' the protostellar sources at each time output using Hyperion, a
parallelized, Monte Carlo radiative transfer dust continuum code [7]. In each case, we observe from 20 different viewing
angles in 10 apertures logarithmically spaced between 1,000 AU and 20,000
AU and at five different grid resolutions. Figure 1b illustrates the SEDs of one
source for four viewing angles observed with a 1000 AU aperture. 

We then compare the observed SEDs sampled at wavelengths appropriate
for the 2MASS, Spitzer, and Bolocam instruments with the 200,000 model
grid published by [8]. This study parametrized the
input properties of the source, disk, envelope and outflow using
14 unique physical variables sampled over a wide
range of apertures and 10 different viewing angles. This work also
provides a simple means to identify models with good-fits to data and
extract the corresponding parameter values. Here, we use these models to
derive the properties for fits satisfying:
$\chi^2-\chi_{\rm best}^2 <3N$,
where $N$ is the number of SED data points and $\chi_{\rm best}$ is
the best-fit model provided $\chi_{\rm best}^2<30N$. Finally, we compare four
derived parameters with the true source properties. More complete
comparisons are presented in [6].
 



\section{Results}
\label{sec:3}

\begin{figure}[!t]
\vspace{-1.2in}
\includegraphics[scale=.36]{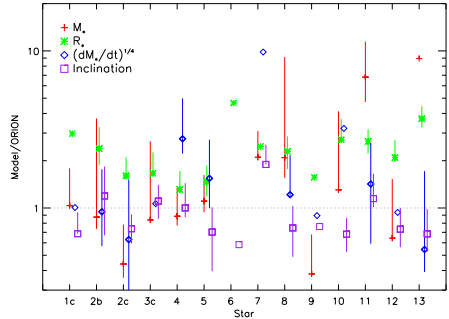}
%
%
\sidecaption
\caption{Ratio of the inferred best model values to the actual 
simulation value for each source, where $M_*$ 
is the protostellar mass, $R_*$ is the protostellar radius, $(dM_*/dt)^{1/4}$ is the 
the accretion rate to the fourth power, and the inclination 
is the tilt with respect to the line of sight (reproduced from [6]). The dotted line 
indicates where the models correctly determine the simulation value.} 
\label{compare}       
\end{figure}

Figure \ref{compare} illustrates how well the good-fit models do in
comparison to the simulated values for protostellar mass,
protostellar radius, accretion rate, and source inclination. We find
that the best-fit models give a reasonable estimate of the
protostellar mass in the cases of more isolated or more massive
protostars (generally those on the left half of Figure \ref{compare}). This
agreement occurs despite the discrepancy between the stellar evolution
models, which causes the models from [8] to systematically overestimate
the protostellar radius by a factor of 2-3. The origin of the
discrepancy is the use of the model tracks by [9] that do not include
accretion. 

The range of inferred protostellar accretion rates typically encompass
the true values, albeit with large spreads. The accretion rates of the
good-fit models often extend over three to four orders of magnitude, which
facilitates general agreement but highlights the difficulty of
precisely constraining the true accretion rate. The source inclination 
proves to be a critical parameter in the comparison; in cases where the
inclination is correctly well-constrained, the other inferred
parameters tend to be more accurate. This suggests that placing limits
on the inclination, e.g., via direct imaging, would improve the
fidelity of the parameter estimation.

\section{Conclusions}
\label{sec:3}

Overall, the comparison between the simulations and analytic models underscores
uncertainties inherent in modeling unresolved observations, especially
in cases where the dust distribution, stellar evolution model, and gas
geometry are not well constrained. Parameters inferred
from the SEDs of unresolved sources should be accepted with caution
and informed by direct imaging wherever possible.

\begin{acknowledgement}
The author thanks the conference organizers for the opportunity to
present this work and acknowledges Thomas Robitaille, Charles Hansen, Christopher
McKee, and Richard Klein, who were collaborators in [6]. Figure \ref{compare} is
reproduced by permission of the AAS. This work
was supported by AST-0901055.
\end{acknowledgement}

\end{document}